\documentclass{aa}  

\usepackage{txfonts}
\usepackage[colorlinks=true, citecolor=blue]{hyperref}
\usepackage{epsf,palatino,url,wrapfig,array,setspace,amsmath,amssymb,fancyhdr,multirow,lscape,appendix,rotating}
\usepackage{graphics,epsfig,txfonts}
\usepackage{graphicx}

\usepackage{longtable}
\usepackage{amssymb}
\usepackage{xcolor}
\usepackage{color}
\usepackage{lipsum}
\usepackage{textcomp}
\usepackage{threeparttable,booktabs}
\usepackage{enumerate}
\usepackage{placeins}
\usepackage{float}
\usepackage[normalem]{ulem}
\usepackage{soul}
\usepackage{blindtext}
\usepackage{grffile}

\usepackage{tikz,hyperref}
\newcommand{\sth}{\sigma_{\rm T}}

\begin{document}

   \title{A study of natural frequencies in a dynamic corona - disk system}

   \author{A. Mastichiadis
          \inst{1}
          \and
          M. Petropoulou\inst{1,2}
          \and 
          N. D. Kylafis\inst{3,4}
          }

    \institute{Department of Physics, National and Kapodistrian University of Athens, University Campus Zografos, GR 15783, Athens, Greece  \\
              \email{amastich@phys.uoa.gr, mpetropo@phys.uoa.gr}
    \and
    Institute of Accelerating Systems \& Applications, University Campus Zografos, Athens, Greece 
    \and 
    University of Crete, Physics Department \& Institute of Theoretical \& Computational Physics, 70013 Heraklion, Greece
    \and
    Institute of Astrophysics, Foundation for Research and Technology - Hellas, 70013 Heraklion, Greece \\
             \email{kylafis@physics.uoc.gr} \\
             }

   \date{Received XXX; accepted XXX}

 
  \abstract
   {
   Black-hole X-ray binaries (BHXRBs) in the hard and hard-intermediate spectral  (and temporal) states exhibit in their power spectra characteristic frequencies called type-C quasi-periodic oscillations (QPOs). Various models that can explain them with various degrees of success have been proposed, but a definitive answer is still missing.
   }
   {
   The hot Comptonizing corona interacting with the cold accretion disk, both of which are central in understanding BHXRBs, is essentially a dynamical system. Our aim is to investigate if the radiative coupling between the two components can produce QPOs. 
   }
   {
   We write and solve the time-dependent equations that describe energy conservation in the system corona - accretion disk.  We examine both constant and variable mass accretion rates.  By necessity, in this first investigation we use a simple model, but it contains all the essential ingredients.
  }
   {
   For a constant mass accretion rate and certain justifiable conditions, the dynamic corona - disk system exhibits oscillations, which die out after a few cycles.  The characteristic frequencies of these oscillations are similar to the ones observed in the power spectra of BHXRBs. For most parameters, the natural frequencies persist even in the case of variable accretion rates.
  }
   {
  We argue that type-C QPOs in BHXRBs could, in principle, arise from the interaction of the hot Comptonizing corona with the much colder accretion disk. If this picture is correct, it has immediate implications for other systems that contain the above constituents, such as active galactic nuclei. 
   }


   \maketitle
%


\section{Introduction}\label{sec:intro}
Black-hole X-ray binaries (BHXRBs) are systems in which mass is transferred from a donor star onto a stellar-mass black hole, converting its gravitational energy into radiation. These systems spend most of their lives in quiescence but occasionally undergo major outbursts that may reach the Eddington limit~\citep[see, e.g.,][for the recent outburst of V404~Cyg]{2017MNRAS.471.1797M, 2019Natur.569..374M, 2020A&A...639A..13K}. Studies of the outburst--quiescence cycle of BHXRBs have revealed dramatic changes in their X-ray spectral characteristics and temporal properties in addition to changes in their luminosity \citep[][]{2006ARA&A..44...49R}.

It is well known that, during their outbursts, BHXRBs exhibit some characteristic spectral and temporal states, which, according to \citet{2010LNP...794...53B}, are:  the quiescent state, the hard state, the hard-intermediate state, the soft-intermediate state (SIMS), and the soft state. A different classification scheme was introduced by \citet{2006ARA&A..44...49R}, and a comparison of the two schemes was presented by \citet{2009MNRAS.400.1603M}. For the purposes of this paper, we will use the \citet{2010LNP...794...53B} classification. 

The standard picture of BHXRBs in the hard and hard-intermediate states involves an accretion disk and a hot corona around (or above) the black hole. The accretion disk emits soft X-ray photons as part of its multi-temperature black-body spectrum. A fraction of the disk soft radiation is directly seen by the observers, while the rest goes into the corona, gets up-scattered by the hot electrons residing therein, and results in a hard X-ray power-law spectrum. Part of the hard photon radiation can be observed directly, while the rest is either absorbed by the disk and re-emitted as soft secondary photons or reflected.  In the hard state, as the name suggests, the X-ray emission is dominated by  the  hard  component, with a very faint thermal disk component sometimes identified.

While the presence of a corona in the hard state is central in describing the phenomenology of BHXRBs, its nature remains largely uncertain.
For example, in some models \citep[for a review, see][]{2007A&ARv..15....1D} the corona is taken to be the inner part of the accretion flow around the black hole, which in the hard and hard-intermediate states is geometrically thick, optically thin, and hot \citep{1994ApJ...428L..13N, 1995ApJ...452..710N}. Alternatively, the corona might be the base of an outflow that emanates from the hot inner flow \citep{2005ApJ...635.1203M} or even the entire outflow, since photons entering 
at its base can travel through the length of the whole outflow  \citep{2003A&A...403L..15R,  2021A&A...646A.112R}.  

A unique feature of BHXRBs is their time variability.
The power spectra of BHXRBs  exhibit characteristic frequencies, called quasi-periodic oscillations \citep[QPOs; for a recent review, see][]{2019NewAR..8501524I}. When BHXRBs are in the so-called hard or hard-intermediate state, where the energy spectrum is dominated by the hard power law, the observed QPOs are called type-C \citep{2005ApJ...629..403C}. In the soft state, where the energy spectrum is dominated by the soft multi-temperature black body emitted from the disk, QPOs are rarely  
detected \citep[see, e.g.,][]{2012MNRAS.427..595M}.  The frequency of type-C QPOs in BHXRBs is in the range of a few tens of millihertz to a few tens of hertz, and it is an increasing function of luminosity \citep[][]{2005A&A...440..207B}. It also correlates with spectral parameters, such as the photon index of the power law \citep{2003A&A...397..729V}. Several scenarios have been proposed for explaining type-C QPOs, including oscillations of boundary layers \citep{2004ApJ...612..988T},
some type of disk instabilities \citep{1999A&A...349.1003T}, or Lense-Thirring precession \citep{1998ApJ...492L..59S,  2009MNRAS.397L.101I}.

The idea that there is an interaction between the corona and the accretion disk was first put forward by \cite{1991ApJ...380L..51H} to explain the X-ray spectra from radio-quiet active galactic nuclei (AGN). Usually, the implicit assumption is that there is a balance between heating and cooling, and the corona is thereby in some stationary state. Even when variations in one or more physical parameters are considered, the changes are assumed to occur in such a way that this balance is not violated. However, there is no {\sl a priori} reason why this should always be the case. This issue could be clarified only with a study of the dynamical behavior of the corona-disk system, and such research, to the best of our knowledge, had not been performed until now.
The aim of the present paper, therefore, is to make an exploratory study of the dynamical aspects of the radiative feedback between the corona and the accretion disk in the hard (and hard-intermediate) state of BHXRBs.

To treat the above nonlinear problem, 
we write two coupled time-dependent equations that describe the evolution of  the energy density of the hot electrons in the corona and the energy density of the hard photons emanating from it. The electrons are cooled due to the Comptonization of soft photons from the disk, including those produced by the reprocessing of the impinging hard photons on the disk.
While these equations are by necessity simplified -- they consider only the integrated energy densities of each population -- they are self-consistent, in the sense that they conserve energy, and they keep the physical essence of the interactions. Thus, for instance,
the hot electrons in the corona are assumed to be heated by  mechanisms that are left unspecified, but are connected with the accretion flow. On the other hand, since the electrons interact with the disk photons, they cool through inverse Compton scattering, and the electron energy losses are taken to be equal to the radiated hard photon luminosity.

The corona-disk system, as described above, has many similarities to the prey-predator dynamic models commonly used in the study of biological and ecological systems \citep[e.g.,][]{Brauer2001}. The prey-predator relationship basically describes the interactions between two species and their mutual effects. For example, in the case of the lynx and the snowshoe hare, when the number of predators (lynx) is low, the preys (hares) thrive.  The population of predators, however, will increase as they can reproduce and survive by feeding on the increasing number of preys. With more lynx hunting, the hare population will rapidly decline, eventually bringing down the population of lynx due to starvation, and the cycle repeats itself. In the corona-disk system under study, hot electrons are the prey and soft disk photons (from the disk and reprocessed) are the predator. When the energy density of the latter is low, the electron energy density builds up in the corona because the inverse Compton scattering cooling is negligible. When the electrons ``thrive,'' the energy density of hard photons impinging on the disk also increases, thus leading to an increase in the reprocessed soft radiation and electron cooling. This, in turn, suppresses the ``growth'' of the electron energy density, eventually leading to a decrease in the electron cooling agents, and the cycle repeats.

Indeed, the solution of the ordinary differential equations that describe the corona-disk coupling shows that the energy densities of the interacting particle species undergo  quasi-periodic changes before they relax to a steady state, that is to say, the  system undergoes damped oscillations. The frequency of the oscillations depends on the mass accretion rate, the radius of the corona, the fraction of hard radiation that is being reprocessed to soft, and the black hole mass. Interestingly, for nominal values of these parameters, the derived frequencies are in the range of type-C QPOs (i.e., from a millihertz to tens of hertz).

This paper is structured as follows. In Sect.~\ref{sec:model} we outline the model and present the equations that describe the interaction between the electrons in the corona and the soft radiation from the disk. In Sect.~\ref{sec:results} we present numerical solutions to the equations of the corona-disk system and investigate the effect of the model parameters on its variability properties. We conclude in Sect.~\ref{sec:discussion} with a discussion of our results.

\section{Model description}\label{sec:model}
We first consider a corona with a thermal electron population of temperature $\Theta = k_{\rm B} T/ m_{\rm e} c^2$ immersed in a low-energy (soft) photon bath with energy density $U_{\rm s}$. The cooling timescale of the electrons due to Comptonization  is
$t_{\rm cool} = m_{\rm e} c^2 (4 \sth c U_{\rm s})^{-1}$ \citep[e.g.,][]{1986rpa..book.....R} and depends solely on the energy density of soft photons. If the electron heating rate is independent of $U_{\rm s}$, as assumed here, it is likely that cooling might not be able to balance heating at all times, and a steady state might not be reached for all conditions.

The irradiation of the disk by the hard coronal photons results in a reflected component and an absorbed one, which is reprocessed. The reprocessed radiation is assumed to be soft, and as such, it also contributes to the cooling of the electrons. Since the reprocessed radiation is originally produced by the electrons, it constitutes a nonlinear cooling term in analogy to the well-known synchrotron self-Compton process for relativistic electrons \citep{1996ApJ...461..657B, 1997A&A...320...19M, 2010A&A...524A..31Z}. 

The study of the nonlinear cooling of the electrons in the corona and its impact on the emergent radiation (hard and reprocessed components) requires a time-dependent approach. In this work we construct a simplified set of equations that describe the interplay between: (i) hot electrons in the corona, which we denote by the subscript  $e$, (ii) hard photons produced through Comptonization, which we denote by the subscript $h$, (iii) soft photons produced from the disk thermal emission, which we denote by the subscript  $ds$, and (iv) soft photons produced via the reprocessing of the irradiating hard component in the disk, which we denote by the subscript $rs$.

Our simple model does not consider the evolution of the energy dependence of the particle densities, only the energy-integrated densities of all components. A similar approach, which has been adopted for the study of coupled nonthermal radiative processes in leptohadronic plasmas, was proven successful at capturing the main dynamical properties of the system \citep{2012MNRAS.421.2325P}.  In the following, we write the equations and relations that describe the temporal evolution of each component. 

\textit{Hot electrons.} We assume that a fixed number of electrons $N_{\rm e}$ resides in the corona, which is modeled as a sphere of radius $R_{\rm c}$. The electrons are heated by some mechanism, which is left unspecified  \citep[for heating models, see, e.g.,][]{1992MNRAS.259..604T, 2000ApJ...534..398M,  2017ApJ...850..141B, 2019MNRAS.484.4920Y, 2021MNRAS.507.5625S}, but it is assumed to depend on the mass accretion rate $\dot{M}$. More specifically, we assume that the energy injection rate to hot electrons is a fraction $f\le 1$ of the rate with which gravitational energy is being released as mass is being accreted from infinity to the radius of the innermost stable orbit (i.e., $f G M_*\dot{M} / R_{\rm isco}$; e.g., \citealt{1991ApJ...380L..51H}). Here, $M_*$ is the mass of the compact object, $R_{\rm isco} = \xi R_{\rm S}$ is the radius of the inner stable circular orbit,  $R_{\rm S}=2 G M_*/c^2$ is the Schwarzschild radius of the compact object, and $\xi=3$ or $1/2$ for nonrotating or maximally (prograde) rotating compact objects, respectively \citep{1949ZhETF..19..951K, 1972ApJ...178..347B}. We note that the specific choice for the heating rate would not alter our main findings as long as it is proportional to $\dot{M}$.

The electron cooling rate (in the limit of small optical depths) can be written as $- N_{\rm e} \Theta m_{\rm e} c^2/t_{\rm cool}$ \citep[e.g.,][]{1986rpa..book.....R}. Then, the equation that describes the energy balance of electrons reads 
    \begin{equation}
    \frac{{\rm d} E_{\rm e}}{{\rm d}t}=f\frac{G M_*\dot{M}}{\xi R_{\rm S}}-4 N_{\rm e}\Theta \sth c U_{\rm s}, 
    \label{eq:electron}
    \end{equation}
where $E_{\rm e} \approx (3/2) N_{\rm e} \Theta m_{\rm e} c^2$ is the total energy content of the electrons in the corona. Because the number of electrons is considered fixed inside the corona, there is no escape term ($-E_{\rm e}/t_{\rm e, esc}$) on the right-hand side of the above equation. We discuss the effects of particle escape in Sect.~\ref{sec:discussion}.

Dividing by the volume of the corona, Eq.~(\ref{eq:electron}) can be written in terms of the thermal electron energy density, $U_{\rm e}$, as
     \begin{equation}
        \frac{{\rm d} U_{\rm e}}{{\rm d} t}=
        \frac{3f}{4\pi}\frac{G M_* \dot{M}}{\xi R_{\rm S} R_{\rm c}^3}-\frac{8}{3}\frac{\sth c}{m_{\rm e} c^2} U_{\rm s} U_{\rm e} .
        \label{eq:electron-density}
    \end{equation}
\textit{Comptonized (hard) photons}. These photons are produced by the hot electrons in the corona. 
Their energy production rate should be equal to the electron loss rate, and their escape from the source is given, in the limit of optically thin coronae, by the source crossing time $t_{\rm c}=R_{\rm c}/c$ \citep[see, e.g.,][]{1986MNRAS.221..931F}. The equation that describes the hard photon energy density can be written as
    \begin{equation}
        \frac{{\rm d} U_{\rm h}}{{\rm d} t}=  \frac{8}{3}\frac{\sth c}{m_{\rm e} c^2} U_{\rm s} U_{\rm e} - \frac{c}{R_{\rm c}}U_{\rm h} .
    \label{eq:comptonized-density}
    \end{equation}
\textit{Disk (soft) photons.} These photons are the soft photons produced at the inner part of the cold accretion disk. For simplicity \citep[see also][]{2021MNRAS.503.5522K}, they are assumed to be produced within a sphere of radius $R_{\rm d}$ by the dissipation of the remaining $1-f$ fraction of the accretion power. In our formalism disk, photons do not depend on the other populations. Therefore, their energy density just outside this sphere can be written directly as 
    \begin{equation}
    U_{\rm ds}=
    \frac{3(1-f)}{4\pi}\frac{G M_*\dot{M}}{\xi R_{\rm S} R_{\rm d}^2c}.
    \label{eq:disc-density}
    \end{equation}
\textit{Reprocessed (soft) photons}. These photons are produced by the hard photons, which illuminate the accretion disk and get absorbed there. Reprocessing in the disk transforms the incident hard radiation to soft.
For simplicity, we neglected the slight temperature increase in the disk  and assumed that the reprocessed photons have the same average energy as the disk photons. Thus, as long as the production rate of the reprocessed soft photons is proportional to the incident rate of hard photons, one can relate their energy densities as
 \begin{equation}
     U_{\rm rs} = a U_{\rm h},
     \label{eq:reflected-density}
 \end{equation}
where $a<1$ is a factor that incorporates both geometrical and albedo effects. Since both the disk and the reprocessed photons are soft, they contribute to electron cooling\footnote{ The reflected hard X-ray photons either escape (and are observed) or go back to the corona. In the second case, their effect on the corona is negligible because the average energy of the reflected photons is slightly lower (due to recoil) than that of the corona photons. Therefore, the reflected photons cannot cool the corona.}
and one can write 
\begin{equation}
     U_{\rm s} = U_{\rm rs} +  f_{\rm dil} U_{\rm ds},
     \label{eq:soft-density}
 \end{equation}
where $f_{\rm dil}=\left(R_{\rm d}/R_{\rm c}\right)^2$, if $R_{\rm d} \le R_{\rm c}$
is a geometric factor that accounts for the dilution of the soft disk photon density in the corona.

It is useful to introduce normalized quantities at this point, namely $\tilde{t} = c t/R_{\rm S}$, $r_{\rm c,d}=R_{\rm c,d}/R_{\rm S}$, and  $\tilde{u}_{\rm i}=4 \pi R_{\rm S}^2 c U_{\rm i}/L_{\rm Edd}$, where $L_{\rm Edd}=4 \pi G M m_{\rm p} c/\sth$ and $\dot{m} = \dot{M}/\dot{M}_{\rm Edd} = \dot{M} c^2/L_{\rm Edd}$. Then, Eqs.~(\ref{eq:electron-density})-(\ref{eq:reflected-density}) can be rewritten in dimensionless form as
\begin{eqnarray}
\label{eq:e-norm}
\frac{{\rm d}\tilde{u}_{\rm e}}{{\rm d}\tilde{t}} & = &   \frac{3f}{2\xi} \frac{\dot{m}}{r_{\rm c}^3} - \frac{4}{3}\frac{m_{\rm p}}{m_{\rm e}} \tilde{u}_{\rm e} \left(\tilde{u}_{\rm rs} + \tilde{u}_{\rm ds} f_{\rm dil}\right), \\ 
\label{eq:corona-norm}
\frac{{\rm d}\tilde{u}_{\rm h}}{{\rm d}\tilde{t}} & = & - \frac{\tilde{u}_{\rm h}}{r_{\rm c}} + \frac{4}{3}\frac{m_{\rm p}}{m_{\rm e}} \tilde{u}_{\rm e} \left(\tilde{u}_{\rm rs} + \tilde{u}_{\rm ds} f_{\rm dil}\right), \\
\label{eq:disc-norm}
\tilde{\rm u}_{\rm ds} & = &  \frac{3(1-f)}{2\xi}\frac{\dot{m}}{r^2_{\rm d}}, \\ 
\label{eq:reflect-norm}
\tilde{\rm u}_{\rm rs} & = & a \tilde{u}_{\rm h}.
\end{eqnarray}

The steady-state energy densities can easily be obtained from Eqs.~(\ref{eq:e-norm})-(\ref{eq:corona-norm}) after setting the time derivatives equal to zero.\ They read 
\begin{eqnarray}
\label{eq:uh-ss-norm}
\tilde{u}_{\rm h}^{\rm (ss)} & = &  \frac{3f}{2\xi} \frac{\dot{m}}{r_{\rm c}^2}, \\ 
\tilde{u}_{\rm e}^{\rm (ss)} &  = &  \frac{3}{4 r_{\rm c}} \frac{m_{\rm e}}{m_{\rm p}}  \frac{f}{1-f + af}
\label{eq:ue-ss-norm}
\end{eqnarray}
or, in dimensional quantities,
\begin{eqnarray}
\label{eq:uh-ss}
U_{\rm h}^{\rm (ss)} & = &  \frac{3f}{\xi}  \frac{\dot{M}c^2}{8\pi c R_{\rm c}^2}, \\ 
U_{\rm e}^{\rm (ss)} &  = &  \frac{3}{8} \frac{m_{\rm e}c^2}{\sth R_{\rm c}} \frac{f}{1-f + af}\cdot 
\label{eq:ue-ss}
\end{eqnarray}
Interestingly, the steady-state energy density of hot electrons does not depend on the mass accretion rate but solely on the size of the corona and $a$ (for $f \approx 1$). Equation~(\ref{eq:ue-ss}) can also be expressed in terms of the electron temperature and Thomson optical depth ($\tau_{\rm T}=n_{\rm e}\sth R_{\rm c}$), leading to  $\tau_{\rm T}\Theta = f/(4(1-f+af))\approx 1/(4a)$ for $f\approx 1$. Thus, for $\tau_{\rm T}\sim 1$ and $a\sim 0.5$, our simple model predicts $k T_{\rm e} \sim 250$~keV. It can also easily be shown that the hard photon luminosity at steady state, $L_{\rm h} = 4 \pi R_{\rm c}^2 c U^{\rm (ss)}_{\rm h}/3$, is equal to the energy dissipation rate at $R_{\rm isco}$ that goes into electron heating in the corona: $f \dot{M} c^2/(2\xi)$. 

\begin{figure*}
    \centering
    \includegraphics[width=0.47\textwidth]{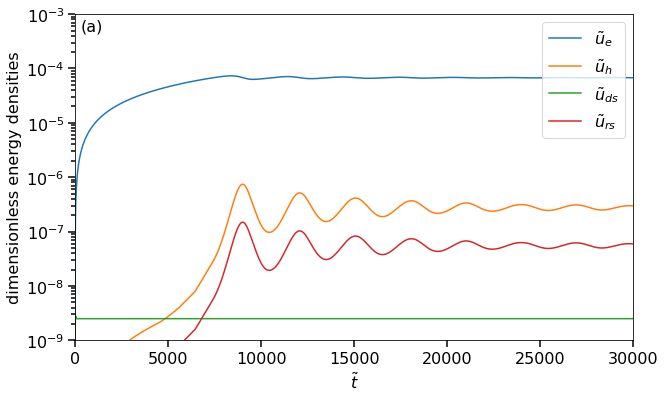}
    \includegraphics[width=0.47
    \textwidth]{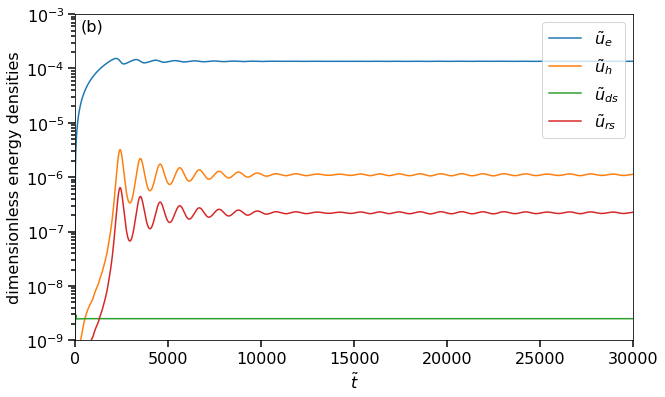}
    \includegraphics[width=0.47
    \textwidth]{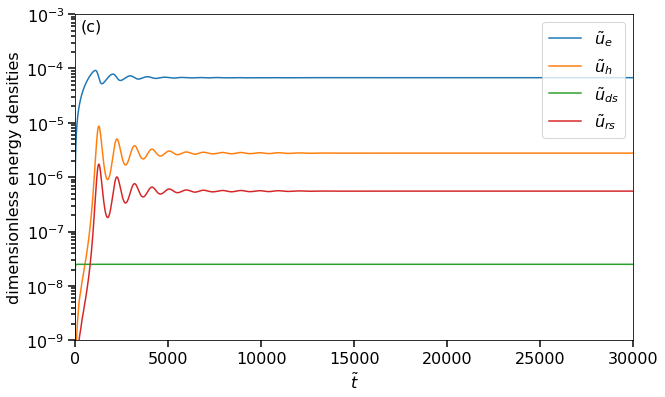}
    \includegraphics[width=0.47
    \textwidth]{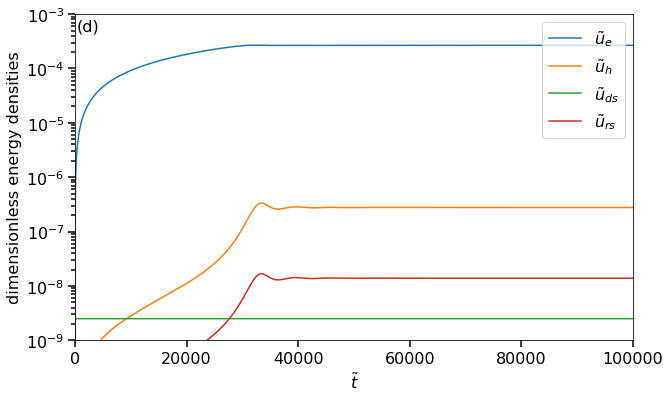}
    \caption{Temporal evolution of energy densities for different model parameters: $\dot{m}=5\times10^{-4}$, $r_{\rm c}=30$, $a=0.2$ (panel a),  $\dot{m}=5\times10^{-4}$, $r_{\rm c}=15$, $a=0.2$ (panel b), $\dot{m}=5\times10^{-3}$, $r_{\rm c}=30$, $a=0.2$ (panel c), and $\dot{m}=5\times10^{-4}$, $r_{\rm c}=30$, $a=0.05$ (panel d). In all cases, $f=0.999, \xi=3, r_{\rm d}=10$,  and time is normalized to $R_{\rm S}/c$.}
    \label{fig:lc-raw}
\end{figure*}

A time-dependent analysis of Eqs.~(\ref{eq:e-norm})-(\ref{eq:corona-norm}) would reveal how different populations reach their steady-state values. This behavior can be qualitatively understood in two limiting cases. If $\tilde{u}_{\rm ds}  \gg \tilde{u}_{\rm rs}$, then Eq.~(\ref{eq:e-norm}) reduces to a linear differential equation of the electron density, since the disk photon density is solely determined by the accretion rate. In this regime, the hot electron energy density reaches a steady state without exhibiting any intrinsic variability. In the case, however, where $\tilde{u}_{\rm ds} \ll \tilde{u}_{\rm rs}$, then Eqs.~(\ref{eq:e-norm}) and (\ref{eq:corona-norm}) constitute a nonlinear system of differential equations, which has, as we will show, some interesting temporal properties.

 \section{Results}\label{sec:results}
\subsection{Constant mass accretion rate}
To better understand the properties of the dynamic corona-disk system under study, we first solved\footnote{We used an implicit multistep method based on a backward differentiation formula for a system of stiff equations.} Eqs.~(\ref{eq:e-norm})-(\ref{eq:corona-norm}), 
assuming that both energy densities are initially very small, namely $\tilde{u}_{\rm e/h}(\tilde{t}=0)=10^{-10}$,
and that the mass accretion rate remains constant. Motivated by observations of accreting systems in the hard state, where the emission is dominated by the hard component, we used $f=0.999$. We verified that our results are not sensitive to the exact value of $f$ as long as it is close to unity (for more details, see later in this section). In what follows, we set $\xi=3$ and $r_{\rm d}=10$, unless otherwise stated.

\begin{figure}
\includegraphics[width=0.47\textwidth]{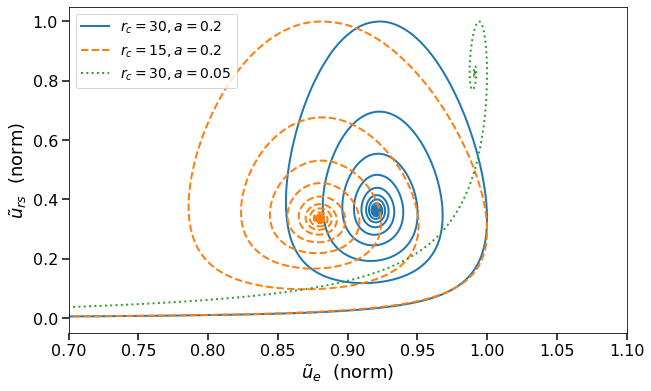}
\caption{Phase-space plot illustrating the dynamic behavior of the corona-disk system. Here, the same parameters as those in panels (a), (b), and (d) of Fig.~\ref{fig:lc-raw} are used. For a better display, the energy densities are normalized to a peak value of one.} 
\label{fig:phase-space}
\end{figure}

\begin{figure*}
    \centering
    \includegraphics[width=0.47\textwidth]{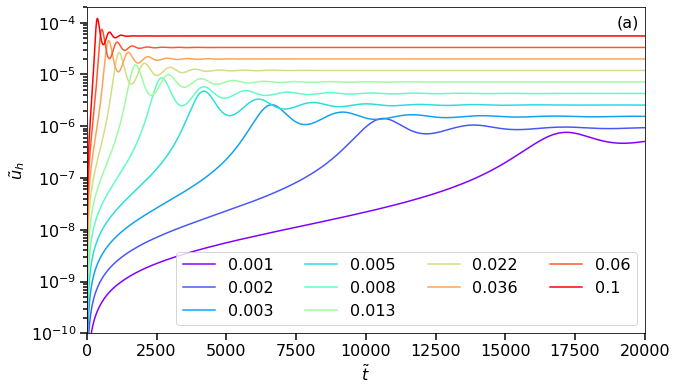}
    \includegraphics[width=0.47\textwidth]{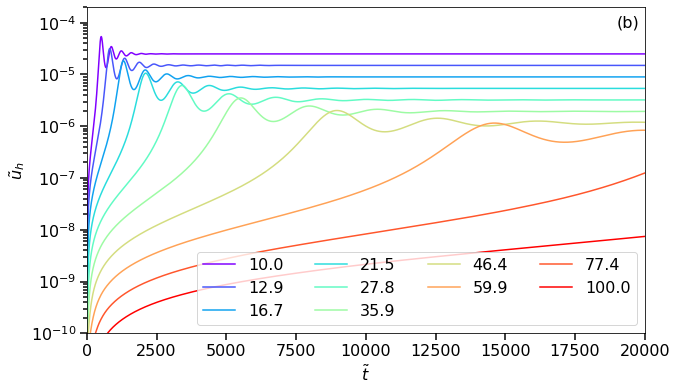}
    \includegraphics[width=0.47\textwidth]{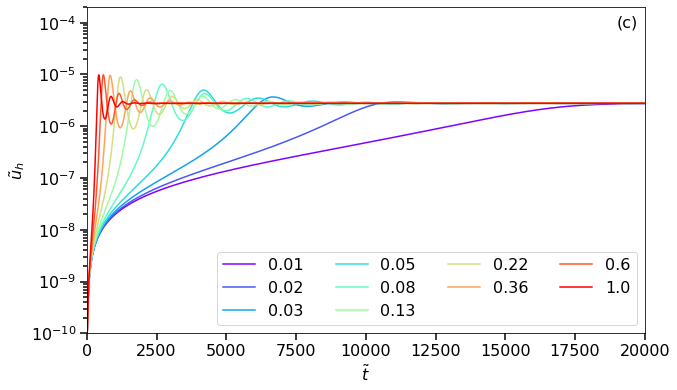}
    \includegraphics[width=0.47\textwidth]{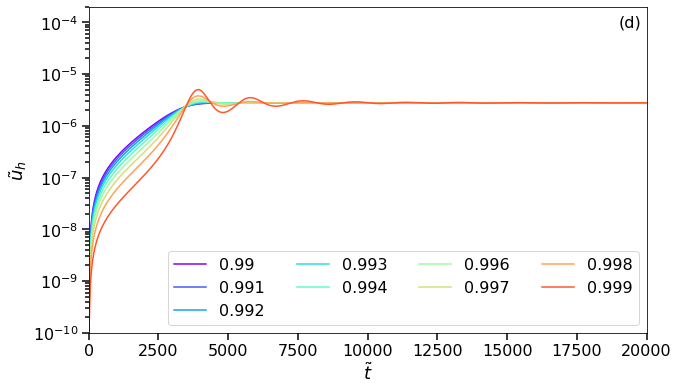}
    \caption{Temporal evolution of the dimensionless hard photon energy density for different values of $\dot{m}$ (panel a), $r_{\rm c}$ (panel b), $a$ (panel c), and $f$ (panel d). Unless stated otherwise, the parameters used are: $\dot{m}=5\times10^{-3}$, $r_{\rm c}=30$, $a=0.05$, and $f=0.999$.  Time is normalized to $R_{\rm S}/c$.}
    \label{fig:lc_param}
\end{figure*}

\begin{figure*}
        \centering
    \includegraphics[width=0.47\textwidth]{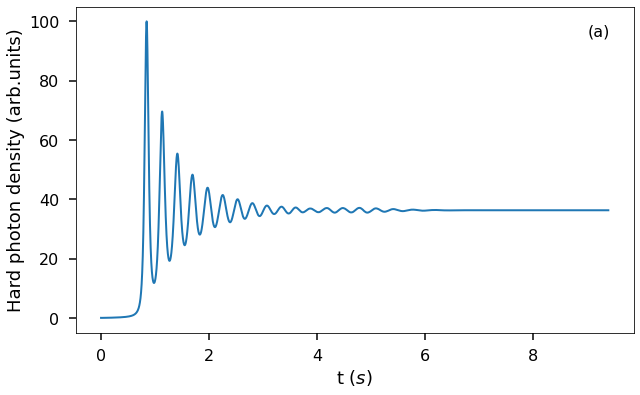}
    \hfill
    \includegraphics[width=0.47
    \textwidth]{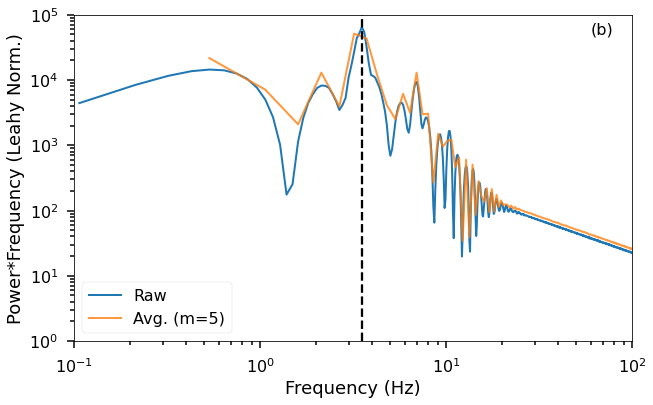}
    \caption{Evolution of the hard photon density in time and frequency domains. Panel (a): Light curve of hard (Comptonized) photons  (rescaled to a maximum value of 100) for the same parameters used in panel (a) of Fig.~\ref{fig:lc-raw}. Panel (b): Power spectrum of the light curve multiplied by frequency using the Leahy normalization. Overplotted in orange is the averaged power spectrum  computed using five segments of the light curve.}
    \label{fig:lc-noise}
\end{figure*}

Figure~\ref{fig:lc-raw} shows the temporal evolution of the dimensionless energy density of the various components for different choices of the accretion rate $\dot{m}$, the corona radius $r_{\rm c}$, and the parameter $a$. The solutions for all components (except the disk soft photons) are, in general,  oscillatory for several cycles before they die out and the system settles into  the steady state described by Eqs.~(\ref{eq:disc-norm}), (\ref{eq:ue-ss-norm}), and (\ref{eq:uh-ss-norm}). The frequency and amplitude  of these transient oscillations depend on the physical parameters, as exemplified in Fig.~\ref{fig:lc-raw}. For instance, an increase in the accretion rate leads to a higher disk photon density, which acts as a stabilizing factor and drives the electrons in the corona into a steady state faster (see the qualitative discussion at the end of the previous section). Similarly, the time window of the transient oscillations decreases for smaller coronae, while the frequency of the oscillations increases (compare panels a and b). As the parameter $a$ decreases, the ratio of the reprocessed soft photon energy density to the hard incident photon energy density decreases. As a result, the cooling of electrons due to the reprocessed soft photon radiation becomes less important, thus weakening the nonlinear coupling between the electron and hard photon populations (see Eqs.~\ref{eq:e-norm} and \ref{eq:corona-norm}). As a result, solutions for lower $a$ values do not exhibit pronounced oscillations, and the system reaches a steady state (compare panels a and d).

Figure~\ref{fig:phase-space} illustrates the prey-predator behavior of the dynamic system under study. Whenever the reprocessed soft photon density increases, the electron energy density decreases because of Compton cooling. Hence, hot electrons represent the prey, and soft photons are the predators. In contrast to the classical prey-predator system, where the oscillations (or limit cycles in phase-space plots) are constant in time, the disk-corona system is always led to a steady state, which is represented by a point in the phase-space plot. Nevertheless, even small perturbations from this steady state lead the system back into an oscillatory mode. We address this point more in the next section and provide analytical arguments in Appendix~\ref{app}.

\begin{figure*}
    \centering
    \includegraphics[width=0.49\textwidth]{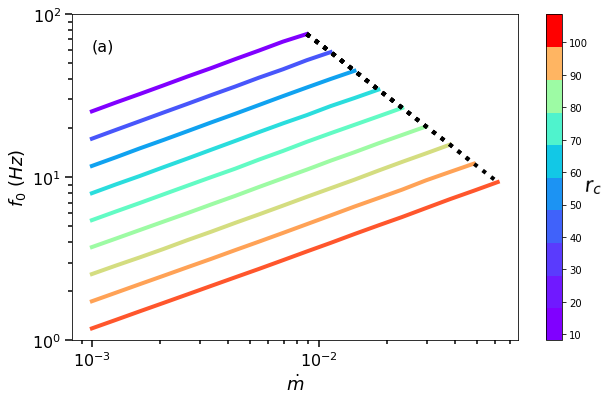}
    \hfill
    \includegraphics[width=0.49\textwidth]{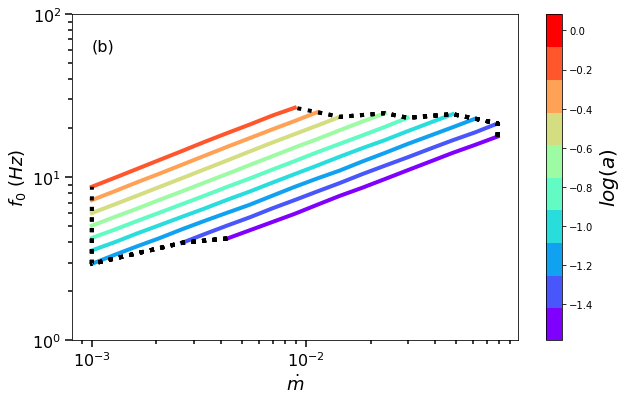}
    \caption{Natural frequency of the dynamic corona-disk system for a nonrotating black hole with mass $M_*=10~M_{\odot}$ as a function of the mass accretion rate (in units of the Eddington rate) for different values of $r_{\rm c}$ (panel a) and  $a$ (panel b), as indicated in the color bar. Beyond (or below) a certain mass  accretion rate, the system reaches a steady state without exhibiting oscillations (dotted black lines). Other parameters used are: $a=0.2$ (panel a) and $r_{\rm c}=30$ (panel b),  with $f=0.999$ in both panels. }
    \label{fig:freq-mdot}
\end{figure*}

Figure~\ref{fig:lc_param} illustrates the dependence of the system's intrinsic variability on $\dot{m}$, $r_{\rm c}$, $a$, and $f$. Here, we display the energy density of hard photons, but similar behavior is also found for the reprocessed soft radiation component and the hot electron population (as previously shown in Fig.~\ref{fig:lc-raw}). The corona-disk system exhibits oscillatory behavior for a wide range of parameter values, except for $f$. In this specific example, we find oscillations only for $f\gtrsim0.997$. However, we typically find that
the system reaches a steady state without signs of intrinsic periodicity if the energy transferred to the corona is $\lesssim 99$ per cent.  Therefore, our simple model predicts no periodicity in soft-state accreting systems where much less power is transferred to the corona. We also notice that, as expected, the steady-state value of the hard photon density does not depend on $a$, while it scales linearly with $\dot{m}$  and $r_{\rm c}^{-2}$ (see also Eq.~\ref{eq:uh-ss-norm}).

Figure~\ref{fig:lc-noise} shows the light curve of hard Comptonized photons (in linear scale and after rescaling to a peak value of 100) for the same parameters used in panel (a) of Fig.~\ref{fig:lc-raw}. For the time units, we assumed a black hole with mass $M_*=10 M_{\odot}$, which corresponds to $R_{\rm S}/c \simeq 9\times10^{-5}$~s. The raw power spectrum of the time series (multiplied by frequency) is displayed in the right panel of the same figure (in blue). The average power spectrum computed after splitting the time series into $m=5$ segments is overplotted (in orange). The peak at $\sim 3.5$~Hz (dashed line) corresponds to the natural frequency of the system (the first harmonic is also visible). This feature resembles a QPO, which is loosely defined as a resolved peak in the power spectrum \citep{1989ARA&A..27..517V}. In reality, the natural frequency may be hidden in the periodogram depending on the noise level and the light curve duration. 

Next we computed the power spectra of noisy light curves, the same way we did for the one displayed in Fig.~\ref{fig:lc-noise}, for different values of the accretion rate, corona radius, and parameter $a$. In all cases, where at least three peaks could be identified in the light curve of hard photons with a peak value of at least 5\% of the asymptotic steady-state value, we determined the natural frequency ($f_0$) of the system. In Fig.~\ref{fig:freq-mdot}, we plot $f_0$ as a function of the dimensionless accretion rate for different values of $r_{\rm c}$ and $a$ (as indicated in the color bars of panels (a) and (b), respectively). The oscillations die out, meaning that the system reaches a steady state after exhibiting one or two bursts above a certain value of the mass accretion rate (as indicated by the dotted black lines). We can see that for low enough values of the $a$ parameter, the oscillations also disappear below a certain $\dot{m}$ value. For this example, in particular, we find no oscillatory solutions for $a\lesssim 0.02$. 
For a black hole with mass $10~M_{\odot}$, as assumed here, we find that the natural frequency of the system lies in the range between a hertz and tens of hertz and depends on the parameters as $f_0 \propto \dot{m}^{1/2} r_{\rm c}^{-3/2} a^{1/2}$. This result is robust -- it does not depend on our method for numerically determining $f_0$. In fact, the above relation can be obtained analytically by performing a stability analysis of Eqs.~(\ref{eq:e-norm})-(\ref{eq:corona-norm}) near the steady state (for more details, see Appendix~\ref{app}). This analysis also reveals the dependence on $\xi$ and $M_*$, which were not varied in the numerical investigation. The full expression for the natural frequency then reads
\begin{eqnarray}
f_0 \simeq 5~{\rm Hz}\
\left( \frac{a}{0.2}\right)^{1/2}\left( \frac{\dot{m}}{10^{-3}} \right)^{1/2}\left(\frac{r_{\rm c}}{30}  \right)^{-3/2}\left( \frac{\xi f}{3} \right)^{-1/2} \left( \frac{M_*}{10~M_{\odot}}\right)^{-1}.
\label{eq:f0}
\end{eqnarray} 

Observations of BHXRBs in the hard and hard-intermediate states indicate that the QPO frequency increases with X-ray luminosity or, equivalently, the mass accretion rate \citep[e.g.,][]{2005A&A...440..207B, 2011MNRAS.411L..66H}. In addition, the corona shrinks in the hard-intermediate state \citep{2018A&A...614L...5K, 2019Natur.565..198K}, also resulting in an increase in the QPO frequency according to Eq.~(\ref{eq:f0}). Finally, as the source evolves in the hard-intermediate state, the inner radius of the accretion disk decreases \citep{2015ApJ...813...84G}, resulting in higher values of the parameter $a$ and higher $f_0$ values. It is intriguing that the dependence of the predicted natural frequency on physical parameters in our simple model agrees qualitatively with observed trends.

\begin{figure*}
    \centering
\includegraphics[width=0.47\textwidth]{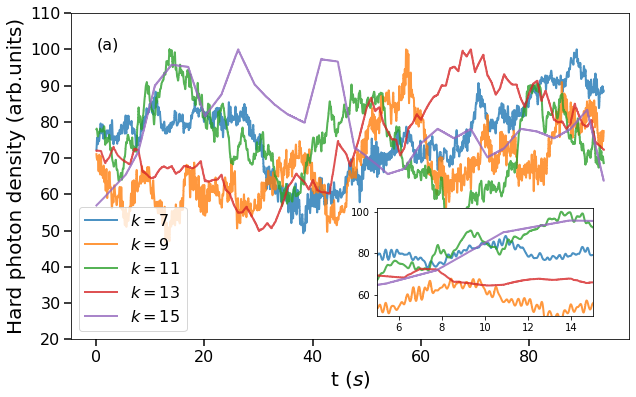}
\hfill
\includegraphics[width=0.47\textwidth]{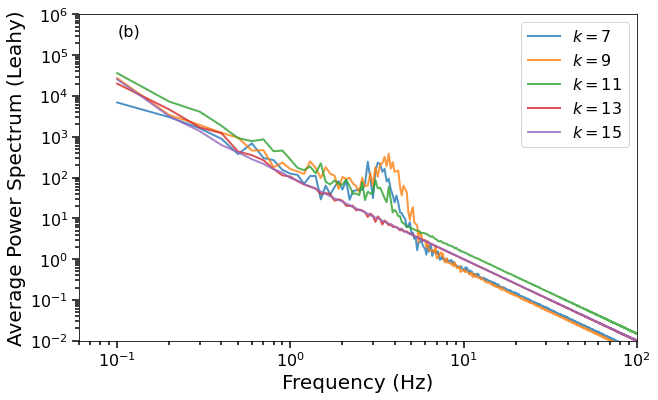}    
    \caption{Hard photon light curves (panel a) and power spectra (panel b) obtained for red-noise mass accretion rates with a minimum to maximum ratio of 2. Other parameters used are: $r_{\rm c}=30$, $a=0.2,$ and $M_*=10 \, M_\odot$.  The initial accretion rate is $\dot m_0=10^{-3}$.}
    \label{fig:mc}
\end{figure*}

\subsection{Variable mass accretion rate}
In the previous section, we demonstrated that the corona-disk system can exhibit quasi-periodicity in the case of a constant accretion rate and under specific initial conditions. To further examine whether these transient oscillations are an intrinsic property of the system, we performed runs in which we allowed the mass accretion rate to vary in time. 

We generated Gaussian-distributed noise with a power-law spectrum ($\propto f^{-2}$) using the Python software {\tt colorednoise}\footnote{\url{https://pypi.org/project/colorednoise/}}, which is based on the algorithm described in \cite{1995A&A...300..707T}. 
The generated red-noise time series, $\{y_i \}$, is then used to compute a time series for the accretion rate as $\dot{m}_i = (y_i+\delta)^{\epsilon}$, for $i=1, \dots, N$. Here, $N=2^{16}$, $\delta$ was chosen to ensure positive values of $\{\dot{m}_i\}$, and $\epsilon$ was chosen such that the maximum and minimum values of the input always differ by a factor of 2. This normalization allowed us to control the dynamic range of the variations in $\dot{m}$ and gives us a basis to compare the output photon light curves. We verified that the power spectrum of $\{\dot{m}_i\}$ is a power law with the same index as the generated time series $\{y_i\}$.

Since the timescale of the mass accretion variations, $\Delta\tau=t_{\rm i+1}-t_{\rm i}$, is not known {\sl a priori}, we studied the system using $\Delta\tau=2^k \, R_{\rm S}/c$ with $k=0,1,2,...$. For any value of $k>0$, we performed a linear interpolation between successive values (i.e., $\dot{m}_{\rm i+1}$ and $\dot{m}_{\rm i}$) and solved  Eqs.~(\ref{eq:e-norm})-(\ref{eq:corona-norm}) with the accompanying relations for the disk and reprocessed soft photon densities. While we started with the same initial conditions as before, here we discarded the first few thousand time steps of the computed light curves in order to remove any dependence on them. 
When $k$ is small, the variations in $\dot{m}$ will be faster than the light-crossing time of the corona, therefore effectively setting a lower limit on $k$. On the other hand,
for large $k$ values changes in $\dot{m}$ happen on a much longer timescale, during which the system will have time to approach a quasi-steady state. For intermediate values of $k$, however, variations in the input will be faster than the time required to reach steady state, and the system should show its characteristic natural frequency, if it exists. 

Our results for different choices of $k$ are presented in Fig.~\ref{fig:mc}. On the left-hand side we show the hard photon light curves (arbitrarily normalized to a peak value of 100), and on the right-hand side we display the corresponding average power spectra. For the averaging of the spectra, we split each light curve into nine segments. As can be seen, a broad feature appears between $\sim 3$ and 4 Hz for $k<11$. For $k=11$ the feature starts diminishing and disappears altogether for larger values of $k$, resulting in
featureless power-law spectra with the same slope as the input time series. Our findings imply that the natural frequency of the system is evident if small-amplitude variations in $\dot{m}$ occur on a timescale $\Delta\tau\simeq (32 - 512) R_{\rm S}/c=(3-50) (M_*/10 M_{\odot})$ ms. 
These values, however, should be taken with some caution. If the changes in $\dot{m}$ are assumed to be discontinuous between successive time steps (i.e., in the form of small-amplitude step functions), then the QPOs persist to much higher values of $\Delta \tau$, yet another manifestation of the strong nonlinearity of the system.

\section{Summary and discussion}\label{sec:discussion}
In this work we have presented a simple model to describe the dynamic coupling of a hot electron population in the corona of accreting objects with the soft radiation emerging from the  disk after the reprocessing of the hard corona radiation. Our approach is simple because we address only the temporal evolution of the system through a set of coupled differential equations,  without treating any spectral features of the radiation. Nonetheless, it is self-consistent, as we couple the hard photon luminosity to the thermal electron losses in the corona.  We find that the temporal properties of the corona-disk system are similar to those of the classical prey-predator dynamical system \citep{1920PNAS....6..410L, 1926Natur.118..558V}, with electrons acting as the prey and soft photons acting as the predator. 

The main physical parameters of the problem are the mass of the central black hole $M_*$, the mass accretion rate (in units of the Eddington rate) $\dot{m}$, the radius of the corona $r_{\rm c}$ (in units of the Schwarzschild radius), and the ratio of hard-to-soft (reprocessed) radiation $a$. Even if all parameters are constant in time, the dynamic corona-disk system exhibits damped oscillations with a characteristic frequency $f_0\propto \dot{m}^{1/2} r_{\rm c}^{-3/2} a^{1/2} M_*^{-1}$. The frequency persists even in cases where $\dot{m}$ varies stochastically and is independent of the initial conditions, strongly indicating that it is an inherent property of the system. The natural frequency of the corona-disk system ranges between $\sim 1$~Hz and tens of hertz for typical parameter values for BHXRBs, with a tendency to increase with increasing $\dot{m}$. Because of the derived scaling of $f_0$ with mass, our model predicts quasi-periodicity on timescales of a few hours ($10^{-5}-10^{-4}$~Hz) to several weeks ($10^{-8}-10^{-7}$~Hz) for supermassive black holes in AGN with masses between $10^6$ and $10^9 M_{\odot}$~\citep[for candidate QPOs in AGN, see, e.g.,][]{2010MNRAS.403....9M, 2018MNRAS.477.3178C, 2021MNRAS.501.5478A}. On the other end, it predicts kilohertz-scale QPOs for neutron stars with masses of $\simeq 1~M_{\odot}$, which indeed seems to be the case for the Rossi X-ray Timing Explorer observations of 4U 1728-34 and Sco~X-1 made more than 25 years ago \citep{1996ApJ...469L...9S, 1996ApJ...469L...1V}.

Disk soft photons tend to stabilize the system by washing out the oscillations and bringing it to a steady state.  We find that the nonlinear coupling of hot electrons and hard photons appears only when the amount of available accretion power goes overwhelmingly to heating up the corona and not the disk (see panel d in Fig.~\ref{fig:lc_param}). Therefore, the recorded oscillations would appear only in the so-called hard and hard-intermediate states of the sources \citep[for a review, see][]{2019NewAR..8501524I}, or, in other words, the system can switch from linear to nonlinear behavior depending on its state. 

We note that, by construction, our model cannot calculate the (dimensionless) temperature $\Theta$ and  the Thomson optical depth of the corona $\tau_{\rm T}$ separately, but rather the product of the two quantities, namely $\tau_{\rm T}\Theta\simeq 0.25\alpha^{-1}$, which is of order unity. The inferred electron temperature generally falls within the range of the one derived from the spectral modeling of BHXRBs in the hard state for $1\lesssim \tau_{\rm T}\lesssim 5$ \citep[see, e.g.,][and references therein]{2012IJMPS...8...73M}.  
A value of $\tau_{\rm T}>1$, however, can delay the escape of photons due to their spatial diffusion, as they scatter off the electrons, and thus our choice of the hard photon escape timescale, $t_{\rm c}=R_{\rm c}/c$ (see Eq.~\ref{eq:corona-norm}), can be considered as a lower limit. In such cases, one should usually introduce a photon diffusion timescale $t_{\rm  c,diff}=R_{\rm c}(1+\tau_{\rm T}/3)/c$ that takes the above effect  into account. Since in most corona models $1\lesssim \tau_{\rm T}\lesssim 5$, introduction of this correction would have modified $f_0$ by a factor of $\sim 2$.

In our model we have only considered inverse Compton scattering as the process coupling photons and electrons. However, there are other competing processes that  contribute to the spectral formation and electron cooling, such as bremsstrahlung and  double Compton scattering. 
We have  examined the effects of the former by adding an energy-loss term in Eq.~(\ref{eq:electron-density}) and a similar source term  in Eq.~(\ref{eq:comptonized-density}) (for rates, see, e.g., \citealt{ 1986rpa..book.....R}),
assuming an electron-proton plasma. 
We have found no significant changes in our results; that is to say, bremsstrahlung, despite its linear nature, cannot stabilize the system. This is to be expected to an extent, since inverse Compton losses are typically much more important as an energy loss mechanism in BHXRBs than bremsstrahlung.

 In our analysis, electrons were confined to the corona.  However, inspection of Eq.~(\ref{eq:electron}) reveals that electron escape can only play a role in cases when the associated escape timescale becomes shorter than the electron cooling timescale.  We typically find that the system quickly reaches a steady state without undergoing oscillations if electrons escape on timescales shorter than $\sim (100-500) R_{\rm c}/c$. 
Therefore, a way of quenching type-C QPO activity is by making electrons escape quickly from the corona. This could occur as the source goes from the hard state to the hard-intermediate one, then to the SIMS, and eventually to the soft state, because the corona shrinks and therefore the outflow becomes continuously narrower.  In the SIMS, the outflow is quite narrow and, for the first time, we see type-B QPOs, which are explained quantitatively as precession of the outflow \citep{2020A&A...640L..16K}.  When the outflow is wide, its precession is not visible; thus, no type-B QPOs are seen in the hard state, only type-C QPOs, which, as we explain, arise from the interaction of the hot corona with the cold disk. Of course, another factor that quenches the type-C QPO activity in the soft state is the increased soft disk radiation, as explained in Sect.~\ref{sec:results}.

We should emphasize that, due to the damped oscillatory behavior of the disk-corona system that we have found, one needs some type of flickering in $\dot m$ in order to sustain the QPOs. Interestingly enough, as we have shown in Sect.~\ref{sec:results}, even small-amplitude perturbations are adequate for that, provided that each perturbation occurs before the oscillations caused by the previous one die out. 
Therefore, perturbations around a constant value of $\dot m$ could produce QPOs of a constant frequency for the whole time the system remains in the hard or hard-intermediate state. We note that such rapid variations (flickering) are evident in the power spectra of BHXRBs.

To explain observed changes in the QPO frequency on longer timescales (e.g., days to weeks), a longer variability timescale needs to be introduced in our model, and this cannot be calculated self-consistently, because it depends on external causes. For instance, 
if the average accretion rate slowly increases on a timescale of weeks, the QPO frequency will also increase.  However, the model still requires the aforementioned flickering in $\dot m$ to sustain the QPOs. We note that since the required flickering is many orders of magnitude faster than the timescale for the hypothesized secular change in $\dot m$, it can safely be assumed that this occurs around the (slowly) changing value of $\dot m$. Clearly,    
no jumps are expected in the QPO frequency evolution unless the flickering in the mass accretion rate changes abruptly by a large factor (see Eq.~\ref{eq:f0}).

Our findings motivate detailed spectro-temporal calculations of the disk-corona system with sophisticated codes that follow the spectral evolution of the interacting particle populations in each time step  \citep[e.g.,][]{2008A&A...491..617B}. Such calculations could also possibly self-consistently address issues such as the possible dependence of $r_{\rm c}$ on $\dot{m}$ and the soft-hard lags. Numerical investigation of feedback processes in nonthermal leptonic and hadronic plasmas~\citep{2012MNRAS.421.2325P, 2020MNRAS.495.2458M} has confirmed the presence of the QPOs found in simplified models. Numerical confirmation of the natural frequency computed here would make the corona-disk radiative feedback a novel scenario for QPOs in BHXRBs.

\begin{acknowledgements}
     We would like to thank the anonymous referee for useful comments that helped us improve the manuscript and Dr. Georgios Vasilopoulos for brainstorming discussions.
      MP acknowledges support from the MERAC Foundation through the project THRILL.
     This research made use of Astropy,\footnote{\url{http://www.astropy.org}} a community-developed core Python package for Astronomy \citep{2013A&A...558A..33A, 2018AJ....156..123A}, and Stingray, a Python library for spectral timing \citep{Stingray2019, Huppenkothen2019}.
\end{acknowledgements}
  
  \bibliographystyle{aa} 
  \bibliography{references} 


\begin{appendix} 
\section{Analytical derivation of the natural frequency}\label{app}
\end{appendix}
Even though finding explicit formal solutions to Eqs.~(\ref{eq:e-norm}) and (\ref{eq:corona-norm})
is not trivial \citep[e.g., for the Lotka-Volterra system, see][]{Evans1999ANT}, one can gain insight into the dynamical properties of the system by performing a stability analysis of its fixed (stationary) points \citep[see also][]{2012MNRAS.421.2325P}.

Equations~(\ref{eq:e-norm}) and (\ref{eq:corona-norm}) can  be recast into the following form,
\begin{eqnarray}
\label{eq:system-1}
\dot{x} & = &  \alpha - \beta x(\gamma y+\delta) \\ 
\dot{y} & = & -y\epsilon  + \beta x(\gamma y+\delta)
\label{eq:system-2}
\end{eqnarray}
where $\dot{x} \equiv {\rm d}x / {\rm d} \tilde{t}$,  $\alpha = 3f\dot{m}/(2 \xi r_{\rm c}^3)$, $\beta=4 m_{\rm p}/ (3 m_{\rm e})$, $\gamma = a$, $\delta = 3(1-f) \dot{m}/(2\xi r_{\rm c}^2)$, and $\epsilon = 1/r_{\rm c}$. 
These equations have one equilibrium point $(x_0, y_0)$ -- commonly referred to as the steady state -- which is given by
\begin{eqnarray}
x_0 & = &  \frac{\alpha}{\beta} \frac{1}{\gamma \frac{\alpha} {\epsilon} + \delta}  \\  
y_0 & = & \frac{\alpha}{\epsilon}.
\end{eqnarray}
We examined the behavior of the system when it is slightly perturbed from its steady state (i.e., $x = x_0 +x'$, $y=y_0+y'$).  We linearized Eqs.~(\ref{eq:system-1})-(\ref{eq:system-2}) with respect to the perturbed quantities $x', y'$ and find
\begin{eqnarray}
\label{eq:perturb}
\begin{pmatrix}
\dot{x}' \\ 
\dot{y}' 
\end{pmatrix} 
= 
\mathbf{J}\left(x_0, y_0\right)
\begin{pmatrix}
x' \\
y'
\end{pmatrix}
,\end{eqnarray}
where 
\begin{equation}
\label{eq:jacobian}
\mathbf{J}\left(x_0, y_0\right) = 
\begin{pmatrix}
-\beta (\gamma y_0 + \delta) & -\beta \gamma x_0 \\ 
+\beta (\gamma y_0 + \delta) & -\epsilon + \beta \gamma x_0 
\end{pmatrix}
\end{equation}
is the Jacobian matrix of the linearized system of equations at the stationary point. 

The evolution of the perturbed quantities near its steady state can be described by the eigenvalues of $\mathbf{J}\left(x_0, y_0\right)$, namely $x', y'\propto e^{\lambda \tilde{t}}$~\citep[e.g.,][]{strogatz:2000}.

To simplify the calculations, we dropped the constant term $\delta$ from the Jacobian matrix since $\delta \ll \gamma y_0$ (or equivalently, $1-f \ll a f$).  
The eigenvalues of $\mathbf{J}\left(x_0, y_0\right)$ are the roots of its characteristic polynomial,
\begin{eqnarray}
P(\lambda) = \lambda^2 + \frac{\alpha \beta \gamma}{\epsilon} \lambda + \alpha \beta \gamma. 
\end{eqnarray}
The system has two complex conjugate eigenvalues,
\begin{eqnarray}
\label{eq:eigenvalues}
\lambda_{1,2} = \frac{- \frac{\alpha \beta \gamma}{\epsilon} \pm i \sqrt{|D|}}{2}
,\end{eqnarray}
if the discriminant $D$ of $P(\lambda)=0$ is negative, which translates to the following constraint
\begin{eqnarray}
\frac{\alpha \beta \gamma} {\epsilon^2} < 4  \Rightarrow \left(\frac{a}{0.1}\right) \left(\frac{\dot{m}}{10^{-3}} \right) \left(\frac{\xi}{1} \right)^{-1} \left(\frac{r_{\rm c}}{10} \right)^{-1} < 10^2.
\label{eq:condition}
\end{eqnarray}

The real part of the eigenvalues is always negative, which means that the steady state is a stable point. In other words, if the system is perturbed from its steady state (e.g., due to a small change in the accretion rate), it will return to it after exhibiting damped oscillations at a frequency given by the imaginary part of the eigenvalues,
\begin{equation}
 \tilde{\omega} = \sqrt{\alpha \beta \gamma} \sqrt{1-\frac{\alpha \beta \gamma} {4 \epsilon^2}} \approx  \left(\frac{2 m_{\rm p}}{m_{\rm e}}\frac{af}{\xi}\frac{\dot{m}}{r_{\rm c}^3}\right)^{1/2}.
 \label{eq:omega}
\end{equation}
The right-hand side of the equation was obtained using the approximation $\alpha \beta \gamma \ll 4 \epsilon^2$, which is satisfied for most parameter values (see also Eq.~\ref{eq:condition}). The natural frequency of the system (in Hz) can then be written as $f_0 = \tilde{\omega} c / (2 \pi R_{\rm S})$ 
\begin{eqnarray}
f_0 \simeq 30~{\rm Hz}\
\left( \frac{a}{0.1}\right)^{1/2}\left( \frac{\dot{m}}{10^{-3}} \right)^{1/2}\left(\frac{r_{\rm c}}{10}  \right)^{-3/2}\left( \frac{\xi f}{1} \right)^{-1/2} \left( \frac{M_*}{10~M_{\odot}}\right)^{-1}.
\label{eq:f0}
\end{eqnarray}

For parameter values that do not satisfy condition (\ref{eq:condition}), the eigenvalues are both real and negative. In this case, the system will return to its steady state (i.e., a stable fixed point) without showing oscillatory behavior. 
\end{document}